# Functional control of anomalous reflection via engineered metagratings without polarization limitations

Jingwen Li,[1,3] Xiao Li,[1,3] Guohao Zhang,[1] Jiaqing Liu,[1] Changdong Chen,[1,*] Youwen Liu,[1,†] and Yangyang Fu,[1,2,‡]

[1]College of Physics, Nanjing University of Aeronautics and Astronautics, and Key Laboratory of Aerospace Information Materials and Physics (NUAA), MIIT, Nanjing 211106, China.
[2]Jiangsu Key Laboratory of Frontier Material Physics and Devices, Soochow University, Suzhou 215006, China.
[3]These authors contributed equally to this work

**Abstract:** Metagratings (MGs) have emerged as a promising platform for manipulating the anomalous propagation of electromagnetic waves. However, traditional methods for designing functional MG-based devices face significant challenges, including complex model structures, time-consuming optimization processes, and specific polarization requirements. In this work, we propose an inverse-design approach to engineer simple MG structures comprising periodic air grooves on a flat metal surface, which can control anomalous reflection without polarization limitations. Through rigorous analytical methods, we derive solutions that achieve perfect retroreflection and perfect specular reflection, thereby leading to functional control over the linearly-polarized electromagnetic waves. Such capabilities enable intriguing functionalities including polarization-dependent retroreflection and polarization-independent retroreflection, as confirmed through full-wave simulations. Our work offers a simple and effective method to control freely electromagnetic waves, with potential applications spanning wavefront engineering, polarization splitting, cloaking technologies, and remote sensing.

## 1 Introduction

Metasurfaces have emerged as a powerful platform for controlling classic waves (e.g., electromagnetic and acoustic waves), facilitating fundamental wave phenomena and advanced device applications[1-6]. In particular, phase gradient metasurfaces[7, 8], featured by periodically arranged supercells that cover a complete phase shift of $2\pi$, have attracted extensive attention in recent years owing to their powerful abilities in arbitrary control of wave propagation. Various adventurous phenomena and promising application have been proposed, such as enhanced spin Hall effect [9, 10], anomalous wave diffraction[11-13], maximum helical dichroism[14], and unidirectional Smith–Purcell radiation[15]. These metasurfaces to achieve anomalous wavefront control rely on the generalized Snell's law[7]. However, when a large impedance mismatch[16] between the incident and reflected or transmitted wave occurs, they suffer from a low conversion efficiency, thus hindering their practical applications.

To improve the conversion efficiency, metagratings (MGs) composed of periodic arrays of meta-atoms have been proposed to break the restriction of continuous phase or impedance profile, thus enabling highly-efficient control on electromagnetic waves[16, 17]. Generally, the electromagnetic response of a MG is collectively determined by the shape, size, and arrangements of meta-atoms in a period. Various types of meta-atoms have been explored to design MGs capable of realizing diverse wavefront functionalities[18-20], such as perfect beam deflection, perfect anomalous diffraction/reflection, broadband and extreme angle imaging. The more functionalities MGs can achieve, the more degrees of freedom are needed in designing meta-atoms. Therefore, the inverse-design methods via optimization algorithms are usually adopted for designing MGs with targeted functionality[21-23]. Although this method can improve both speed and accuracy, but these complex structures together with extreme parameters require time-consuming optimization processes and precise fabrication techniques. Recent studies have explored MGs with simple structures[24-26], but the majority remain designed for a single polarization, often transverse magnetic (TM) polarization, with limited investigation into transverse electric (TE) polarization. Therefore, there is a critical need to develop a straightforward inverse design methodology for MGs that allows for the control of electromagnetic waves without polarization limitations. This advancement holds significant promise for achieving exciting wave phenomena and advanced device applications.

---

[*]cdchen@nuaa.edu.cn
[†]ywliu@nuaa.edu.cn
[‡]yyfu@nuaa.edu.cn

In this work, we introduce an inverse-design strategy for engineering MGs composed of periodic air grooves on a metallic film, aimed at manipulating anomalous reflection phenomena in both TM and TE wave scenarios. Through rigorous analysis and numerical simulations, we derive analytical formulas enabling the achievement of perfect retroreflection and specular reflection, facilitating versatile control over linearly-polarized electromagnetic waves. By adjusting the groove depth via our inverse-design method, we demonstrate the ability to achieve polarization-dependent retroreflection and polarization-independent retroreflection. Furthermore, these effects are observed across various incident wavelengths, showcasing the broad applicability of our approach in many applications[27-31], including wavefront control, polarization manipulation, carpet cloaks, and remote sensing.

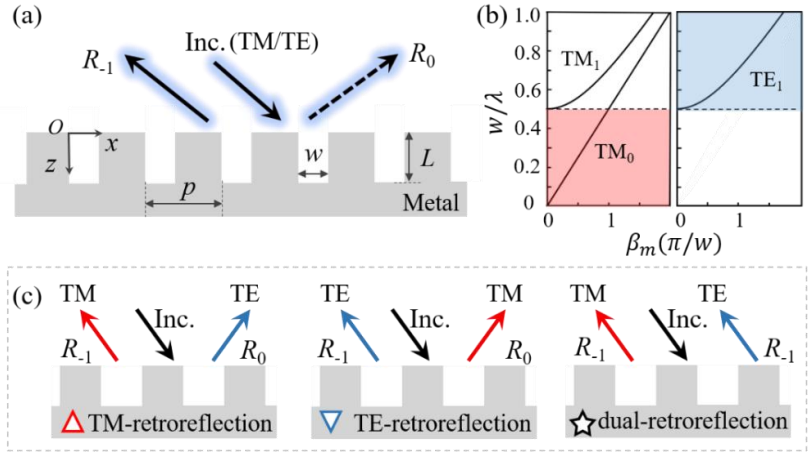

**Fig. 1**. (a) Schematic diagram of retroreflection effect for incident TM and TE waves in inversely-designed MGs, which is composed of periodic air grooves. (b) The dispersion relationship between propagation constants of guided modes and the groove width for TM and TE waves, and the red and blue region indicate that the field in the groove is only composed of the propagating fundamental guided mode. (c) Schematic diagram of polarization-dependent retroreflection (left and middle), dual-polarization retroreflection (right).

## 2 Theory and model

The proposed MG structure is shown in Fig. 1(a), and it is a two-dimensional metal film drilled with periodic air grooves. The MG period is $p$, and the width and the depth of the groove are $w$ and $L$, respectively. We consider the linearly-polarized waves, i.e., TE (the electric field along y direction) or TM (the magnetic field along y direction) waves, incident on the MG. In general, the field distribution in the air groove is composed of all guided modes. For the groove width smaller than a half-wavelength

(see the red region in Fig. 1(b)), it is only the fundamental guided mode of TM waves in the groove, while there are no guided modes for TE waves due to the cut-off effect. In this case, anomalous reflection of TM waves is realized by inversely designing geometric parameters of the groove, which leads to the polarization-dependent retroreflection (see left plot of Fig.1(c)). As the groove width increases ($0.5\lambda < w < \lambda$), higher-order guided modes of TM waves exist in the groove, while it is only the fundamental guided mode of TE waves in the groove (see the blue region in Fig. 1(b)). Compared with TM waves, it is easy to control anomalous reflection of TE waves by inversely designing the MG. In this case, anomalous retroreflection of both TM and TE waves could be controlled in a versatile way to realize these wave functionalities of polarization-dependent retroreflection, polarization-independent retroreflection, and the associated polarization control (see Fig. 1(c)).

Next, we will exhibit the inverse-design process of the MG, which can realize perfect anomalous reflection and polarization control. We first study the case of $w \leq 0.5\lambda$. By considering the incident TM waves, the proposed MG is designed to achieve anomalous reflection with a retroreflection effect[24, 25]. As the MG is a periodic structure, the incident and reflected waves are governed by,

$$k_0 \sin\theta_r + k_0 \sin\theta_i = 2\pi n/p, \tag{1}$$

where $k_0 = 2\pi/\lambda$ is the wave vector in vacuum space, $n$ is the diffraction order, $\theta_i$ and $\theta_r$ are the incident and reflected angle. The MG is engineered to suppress undesired diffraction orders and reroute the incident power towards the desired one with unity efficiency. To achieve the retroreflection effect of $n = -1$ in the MG, the relationship between the period $p$, the incident angle $\theta_i$ and the wave vector $k_0$ should satisfy $p = \pi/k_0 \sin\theta_i$. When the incident angle is larger than approximately 20°, only two propagating diffraction orders of $n = 0$ and $n = -1$ can exist. That is to say, when the width and depth of the groove are well designed, the specular reflection of the $n = 0$ order could be suppressed to achieve perfect retroreflection of the $n = -1$ order. By employing the diffraction grating theory, we can obtain the reflection coefficients of different diffraction orders of the MG. In fact, a simple inverse-design expression cannot be obtained if considering all guided modes in the groove. Previous work has revealed that higher-order evanescent guided modes in the groove can contribute to the retroreflection efficiency[32]. To achieve the inverse design, we consider the fundamental mode ($m = 0$) and the first-order evanescent guided mode ($m = 1$) in the air groove. An analytical solution for perfect retroreflection is obtained by solving $R_0 = 0$ (see Note 1 in the supplementary information),

$$\frac{w^2}{2}\cot(\beta_0 L)\cot(\beta_1 L) + C_1 w \cot(\beta_0 L)\beta_1 + C_2 w \cot(\beta_1 L)\beta_0 = C_3 \beta_0 \beta_1, \tag{2}$$

where $C_1$, $C_2$ and $C_3$ are known constants related to the coupling coefficient, $\beta_0$ and $\beta_1$ represent the propagation constants of the first and second TM guided modes in the groove, respectively. Based on Eq. (2), we can design the MG to realize perfect retroreflection for TM waves.

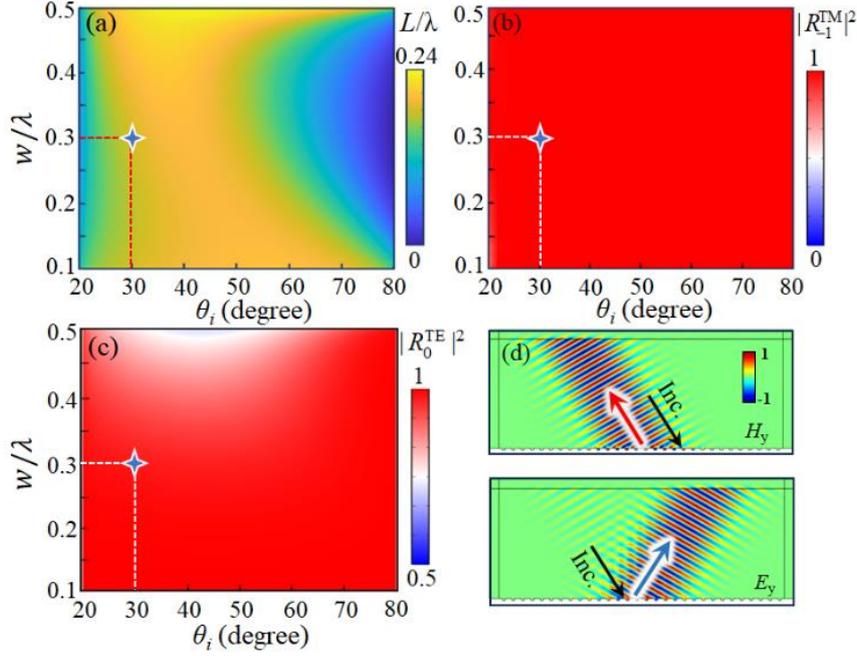

**Fig. 2**. (a) The analytical solution of the groove depth ($L$) and width ($w$) for achieving perfect TM retroreflection based on Eq. (2). Numerically calculated efficiencies of TM-retroreflection (b) and TE specular reflection (c) using the analytical solution in (a). (d) Simulated scattered field patterns ($H_y$ and $E_y$) for TM and TE Gaussian beam incident on the MG, respectively, in which $w = 0.3\lambda, L = 0.2\lambda$, $\theta_i = 30°$ and the operating frequency is 2 GHz.

To verify the inverse-design method, we analytically show the groove depth of the MG for achieving perfect retroreflection of TM waves by varying the groove width and the incident angle, as shown in Fig. 2(a), where the working frequency is 2 GHz. We further perform numerical simulations using COMSOL Multiphysics to calculate the retroreflection efficiency of TM waves by employing the analytical solution in Fig. 2(a). As shown in Fig. 2(b), the unity retroreflection efficiency ($|R_{TM}^{-1}|^2$) for TM waves exists almost in the whole parameter space, which examines the validity of the Eq. (2). We also numerically study the specular reflection ($|R_{TE}^{0}|^2$) of TE waves by adopting the analytical solution in Fig. 2(a), and nearly-perfect specular reflection is seen in the whole parameter space (see Fig. 2(c)) owing to the cutoff effect of the narrow groove width. These results show that the designed MG can realize perfect retroreflection for TM waves and perfect specular reflection for TE waves. For example, when the incident angle is $\theta_i = 30°$, the MG parameters of realizing perfect TM retroreflection are $w = 0.3\ \lambda$ and $L = 0.2\ \lambda$ (the blue stars in Fig. 2). By considering TM and TE Gaussian beam incident on the MG, the corresponding scattering field patterns are shown in Fig. 2(d). We see that the incident TM beam is perfectly retroreflected, while it is nearly-unity specular reflection for the incident TE beam. Thus, such polarization-dependent retroreflection can be used for realizing a polarization splitter.

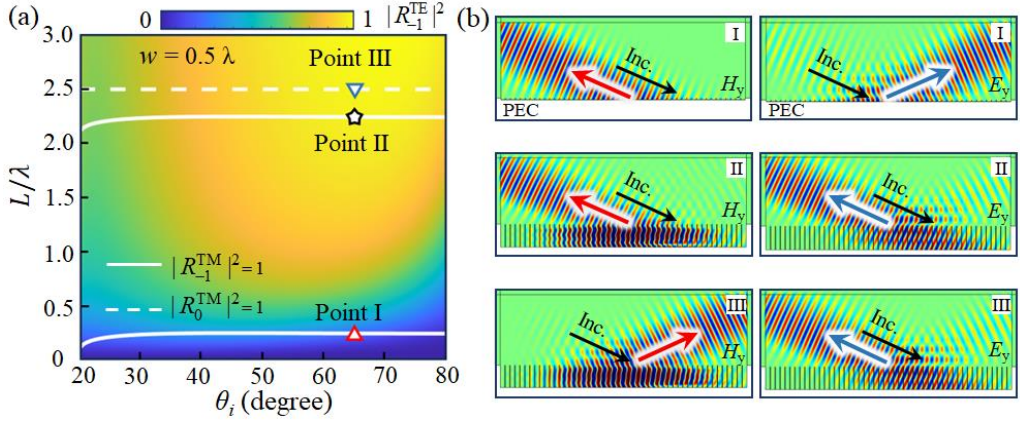

**Fig. 3**. (a) Numerically calculated reflection efficiency of the MGs versus the incident angle and the groove depth, in which $w = 0.5\lambda$ and the working frequency is 2 GHz. The curve in (a) is the analytical solution for perfect retroreflection, and the dotted line is the analytical solution for perfect specular reflection for TM polarized waves. (b) Simulated scattered field patterns ($H_y$ and $E_y$) for a Gaussian beam incident on the designed MGs with different groove depth, respectively, in which $w = 0.5\lambda$, $\theta_i = 65°$. (I) $L = 0.25\lambda$; (II) $L = 2.25\lambda$; (III) $L = 2.5\lambda$.

In above case, $w = 0.5\lambda$ is a critical state, as the propagation constant of the first-order guided mode of TM waves in the groove is exactly zero and only the fundamental mode exists inside the air groove. Therefore, Eq. (2) can be simplified to the following formula (see Note 1 in the supplementary information),

$$\cot(\beta_0 L) = \frac{i\beta_0 w}{p}\sum_{n\neq 0,-1}\frac{\text{sinc}^2(k_z^n w/2)}{k_z^n}, \quad (3)$$

where $k_{x,n} = k_0 \sin\theta_i + 2\pi n/p$ is the wave vector along the $x$-direction, $k_{z,n} = \sqrt{(k_0)^2 - (k_{x,n})^2}$ is the wave vector along the $z$-direction. In fact, Eq. (3) is the analytical solution for perfect retroreflection by only considering the fundamental mode in the groove. Thus, it can be also used to design perfect retroreflection of TM waves except for a larger incident angle and a wider groove width (see Fig. S1 in supplementary information). Moreover, as seen from Fig. 2(c), the incident TE waves can couple with the MG when the groove width is around $w = 0.5\lambda$, which can lead to the generation of retroreflection. To further reveal the retroreflection performance, we numerically study the reflection efficiency of $n = -1$ order for TE waves varying with the incident angle and the groove depth at $w = 0.5\lambda$ (see Fig. 3(a)). We can see that nearly-unity TE retroreflection can appear for a larger incident angle and a deeper groove. Based on Eq. (3), we also show the analytical solutions of perfect TM retroreflection ($|R_{-1}^{TM}|^2 = 1$) and its specular reflection ($|R_0^{TM}|^2 = 1$, see Note 1 in the supplementary information) in Fig. 3(a), as indicated by the white solid and dashed curves, respectively. By observing these results in Fig. 3(a), we can control the retroreflection of both TM and TE waves and realize the polarization control by changing the groove depth. For example, we use $\theta_i = 65°$ to study the wave functionalities. By increasing the groove depth from Point I to Point III in Fig. 3(a), the MG can achieve the transformation from TM-retroreflection to dual retroreflection and TE-retroreflection. Full-wave simulations in Fig. 3(b) show these retroreflection phenomena of the designed MGs, which demonstrate its ability in realizing diverse retroreflection effects. By employing the inverse-design method, anomalous reflection and polarization control can be realized in the MG with $w \leq 0.5\lambda$.

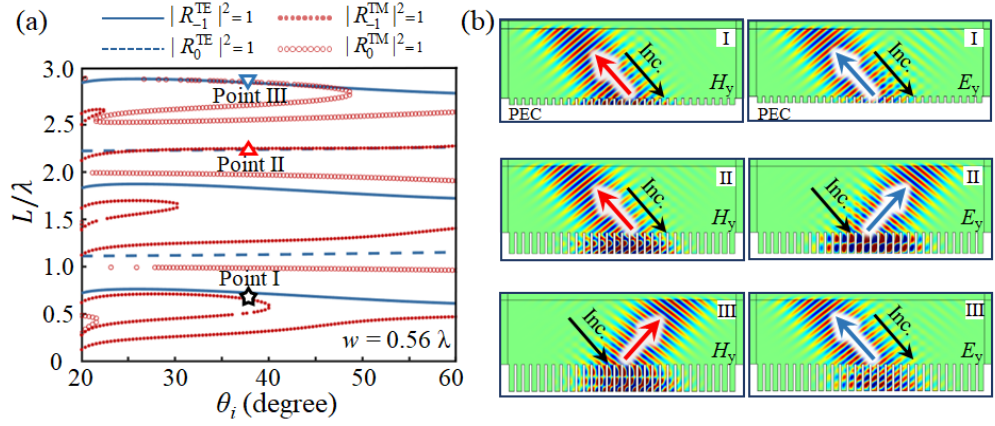

**Fig. 4.** (a) The analytical solutions for perfect retroreflection (blue curve for TE, red scatter dots for TM) and perfect specular reflection (blue dashed for TE, red hollow dots for TM), in which $w = 0.56\lambda$ and the working frequency is 2 GHz. (b) Simulated scattered field patterns ($H_y$ and $E_y$) for a Gaussian beam incident on the designed MGs with different groove depth, respectively, in which $w = 0.56\lambda$, $\theta_i = 38°$. (I) $L = 0.67\lambda$; (II) $L = 2.25\lambda$; (III) $L = 2.95\lambda$.

As mentioned above, for the groove width at $0.5\lambda < w \leq \lambda$, TE waves can be chosen to study the inverse-design method for manipulating anomalous reflection. Similar to the analytical process of the TM case, we consider the fundamental mode ($m = 1$) and the first-order evanescent guided mode ($m = 2$) of TE waves in the air groove to obtain the analytical solution for achieving its perfect retroreflection (see Note 2 in the supplementary information),

$$\frac{w^2}{4}\beta_1\beta_2\cot(\beta_1 L)\cot(\beta_2 L) - F_1\frac{w}{2}\beta_1\cot(\beta_1 L) - F_2\frac{w}{2}\beta_2\cot(\beta_2 L) = F_3, \qquad (4)$$

where $F_1$, $F_2$ and $F_3$ are known constants related to the coupling coefficients, $\beta_1$ and $\beta_2$ represent the propagation constants of the first and second TE guided modes, respectively. However, as there is a restriction between the width and the period of the air groove ($w < p$), so the analytical solution for TE retroreflection is limited by $w < \pi/k_0\sin\theta_i$, i.e., it has a limited parameter space for realizing TE retroreflection. By only considering the first TE guided mode in the groove, the analytical solution for achieving perfect TE retroreflection is written as (see Note 2 in the supplementary information),

$$\frac{w}{2}\beta_1\cot(\beta_1 L) = \frac{2i}{p}\left(\frac{\pi}{w}\right)^2 \sum_{n\neq 0,-1} k_z^n \frac{1+\cos(k_x^n w)}{\left(\left(\frac{\pi}{w}\right)^2 - (k_x^n)^2\right)^2}. \qquad (5)$$

By employing the analytical solutions of Eq. (4) and Eq. (5), numerical simulations are conducted to demonstrate the unity reflection efficiency of $n = -1$ (see Fig. S2 in supplementary information) in both cases, which implies that the contribution of the evanescent waves could be neglected. Therefore, Eq. (5) is much simple to design perfect TE retroreflection. Moreover, we also deduce the analytical equation for realizing perfect TE specular reflection ($|R_0^{TE}|^2 = 1$), shown in Note 2 in the supplementary information. Accordingly, four analytical equations are obtained to control perfect retroreflection/specular reflection for both TM and TE waves, which is useful to inversely design MGs with desired functionalities. For example, when the groove width is $w = 0.56\lambda$, the TE retroreflection can happen for $\theta_i \in (20°, 60°)$. Here, based on these inverse-design equations, we show the analytical solutions of perfect retroreflection and perfect specular reflection for TM waves ($|R_{-1}^{TM}|^2 = 1$ based on Eq. (2), $|R_0^{TM}|^2 = 1$ based on Eq. (S1-25) in supplementary information) and TE waves ($|R_{-1}^{TE}|^2 = 1$ based on Eq. (5), $|R_0^{TE}|^2 = 1$ based on (S2-25) in supplementary information) in Fig. 4(a). Besides, Eq. (2) and Eq. (S1-25) are non-periodic and disorderly, which results in the not continuous and discretized TM solutions in Fig. 4(a). By observing these results in Fig. 4(a), we can inversely design the MG to achieve the transformation of wave functionalities from dual retroreflection (Point I) to TM-retroreflection (Point II) or TE-retroreflection (Point III) by increasing the groove depth at $\theta_i = 38°$. These anomalous reflection phenomena are revealed by full-wave simulations in Fig. 4(b). By choosing other parameters of the groove width, the polarization-dependent retroreflection and dual-polarization retroreflection can happen at either the same incident angle or different incident angles (see Fig. S3 and Fig. S4 in the supplementary information).

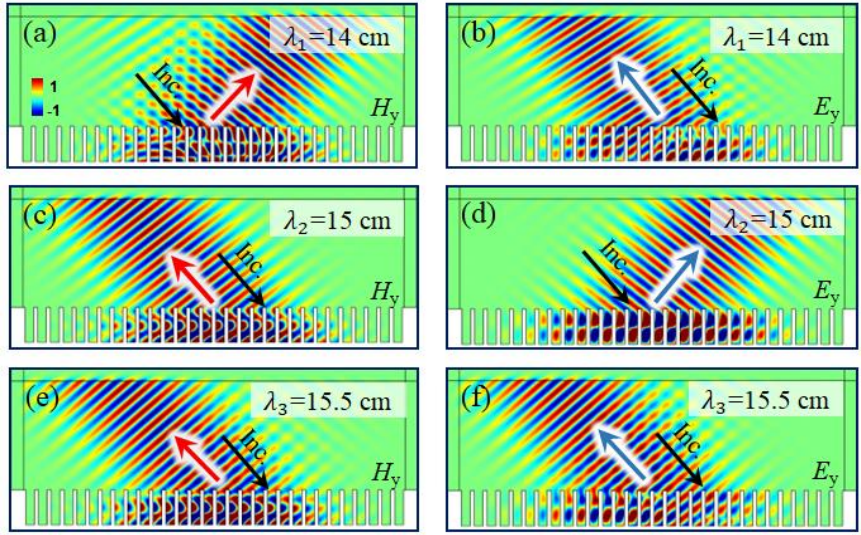

**Fig. 5**. Simulated scattered field patterns ($H_y$ and $E_y$) for a Gaussian beam with $\theta_i = 38°$ incident on the designed MG with $w = 8.4$ cm, $L = 22.5$ cm and $p = 12.2$ cm for different operating wavelengths.

In above discussion, these anomalous reflection phenomena in a diverse manner are controlled by changing the groove depth at a specific wavelength, which hinders their applications across different scenarios. Therefore, we investigate the frequency response of the MG with a fixed structure. By inversely designing a particular MG (Point II in Fig. 4(a)), i.e., $w = 8.4$ cm, $L = 22.5$ cm and $p = 12.2$ cm, TM-retroreflection of $\theta_i = 38°$ is realized at the operating wavelength $\lambda = 15$ cm. By numerically study the reflection efficiency of $n = -1$ in such a MG versus the operating wavelength and the incident angle for both TM and TE waves, we can find that there are some special wavelengths to realize different phenomena of anomalous reflection in a diverse manner (see Fig. S5 in the supplementary information). For instance, when $\theta_i = 38°$, we can achieve the transformation of the wave phenomena from TE-retroreflection to TM-retroreflection or dual retroreflection by only changing the working wavelength, as numerically demonstrated in Fig. 5. For the wavelength at $\lambda = 15$ cm, the incident TM Gaussian beam is perfectly retroreflected (see Fig. 5(c)) and the mirror reflection of the incident TE Gaussian beam is seen in Fig. 5(d). When the operating wavelength decreases to 14 cm, the TM beam is perfectly reflected in a specular way (see Fig. 5(a)), while for the incident TE Gaussian beam, the quasi-retroreflection can happen with an angle deviation of 4° from the exact retroreflection, as shown in Fig. 5(b). When the operating wavelength increases to 15.5 cm, the quasi-retroreflection with an angle deviation of 4° can happen for both TE and TM Gaussian beams, as shown in Figs. 5(e) and 5(f). This deviation arises because the reflection angle varies with the incident wavelength while maintaining a fixed period $p$. Therefore, by inversely designing a particular MG, the functional control over anomalous reflection can be realized by only changing the working wavelength.

## 3 Conclusion

In summary, we theoretically propose and numerically demonstrate a simple MG structure to achieve anomalous reflection and polarization control for both TM and TE waves in an inverse manner. We present specific analytical formulas tailored for designing MGs capable of achieving various wave functionalities, such as polarization-dependent retroreflectors, polarization-independent retroreflectors, and polarization splitters, which may find applications in wavefront engineering, polarization splitting, cloaking technologies, and remote sensing. Our work offers an efficient and convenient approach for freely controlling anomalous reflection with perfect efficiency, and demonstrates the potential for some device applications in the microwave or terahertz range. For example, by involving the inverse-design without polarization limitations for metgratings made of vanadium dioxide[26], one can achieve functional switch between perfect retroreflection and high-efficiency absorber, which can enrich and develop device applications at terahertz frequencies.

**Acknowledgements**
This work was supported by the National Natural Science Foundation of China (12274225), Natural Science Foundation of Jiangsu Province (BK20230089), and Fundamental Research Funds for the Central Universities (NE2022007, NS2023056).